\documentclass[12pt]{article}

\usepackage{scicite}

\usepackage{times}

\usepackage[final]{graphicx}     
\graphicspath{{figures/}}                 

\usepackage{amsmath}
\usepackage{amssymb}
\usepackage{float}
\usepackage{subfig}

\newenvironment{sciabstract}{%
\begin{quote} \bf}
{\end{quote}}

\newcommand\spinup{|\!\!\uparrow\rangle}
\newcommand\spindown{|\!\!\downarrow\rangle}

\newcounter{lastnote}
\newenvironment{scilastnote}{%
\setcounter{lastnote}{\value{enumiv}}%
\addtocounter{lastnote}{+1}%
\begin{list}%
{\arabic{lastnote}.}
{\setlength{\leftmargin}{.22in}}
{\setlength{\labelsep}{.5em}}}
{\end{list}}

\title{Metallic and Insulating Phases of Repulsively Interacting Fermions in a 3D Optical Lattice}

\author
{U. Schneider$^{1}$, L. Hackerm\"uller$^{1}$, S. Will$^{1}$, Th. Best$^{1}$, and I. Bloch$^{1}$,\\ 
T.~A. Costi,$^{2}$ R.~W. Helmes$^{3}$, D. Rasch$^{3}$,  and A. Rosch$^{3}$\\ \\
\normalsize{$^{1}$Institut f\"ur Physik, Johannes Gutenberg-University, 55099 Mainz, Germany}\\
\normalsize{$^{2}$Institut f\"ur Festk\"orperforschung and Institute for Advanced Simulation,}\\ \normalsize{Forschungszentrum J\"ulich, 52425 J\"ulich, Germany}\\
\normalsize{$^{3}$Institut f\"ur Theoretische Physik, University of Cologne, 50937 Cologne, Germany}\\
\\
\normalsize{$^\ast$To whom correspondence should be addressed; E-mail:  bloch@uni-mainz.de.}\\
}
\date{}

\begin{document} 


\baselineskip24pt


\maketitle


\begin{sciabstract}
The fermionic Hubbard model plays a fundamental role in the description 
of strongly correlated materials. 
Here we report on the realization of this Hamiltonian using a repulsively 
interacting spin mixture of ultracold $^{40}$K atoms in a 3D optical lattice. 
We have implemented a new method to directly measure the compressibility of the 
quantum gas in the trap using in-situ imaging and 
independent control of external confinement and lattice depth. 
Together with a comparison to ab-initio Dynamical Mean Field Theory
calculations, we show how the system evolves for increasing confinement 
from a compressible dilute metal over a strongly-interacting Fermi liquid 
into a band insulating state. For strong interactions, we find evidence for an 
emergent incompressible Mott insulating phase.  
\end{sciabstract}

\paragraph*{Introduction.}

Interacting fermions in periodic potentials lie at the heart of modern condensed matter physics, presenting some of the most challenging problems to quantum many-body theory. A prominent example is high-T$_c$ superconductivity in cuprate compounds~\cite{Lee:2006} and the recently discovered iron-arsenic alloys~\cite{Kamihara:2008}. In order to capture the essential physics of such systems, the fermionic Hubbard Hamiltonian~\cite{Hubbard:1963a} has been introduced as a fundamental model describing interacting electrons in a periodic potential~\cite{Lee:2006,GebhardBook}. In a real solid, however, the effects of interest are typically complicated by e.g. multiple bands and orbital degrees of freedom, impurities and the long-range nature of Coulomb interactions which becomes especially relevant close to a metal to insulator transition. It is therefore crucial to probe this fundamental model Hamiltonian in a controllable and clean experimental setting. Ultracold atoms in optical lattices provide such a defect-free system~\cite{Jaksch:2005,Bloch:2008c}, in which the relevant parameters can be independently controlled, allowing quantitative comparisons of the experiment with modern quantum many-body theories. For the case of bosonic particles~\cite{Fisher:1989,Jaksch:1998}, a series of experiments carried out in the regime of the superfluid to Mott insulator transition~\cite{Greiner:2002a,Stoferle:2004,Spielman:2007} have demonstrated the versatility of ultracold quantum gases in this respect. For both bosonic and fermionic systems, the entrance into a Mott insulating state is signaled by a vanishing compressibility, which can in principle be probed experimentally by testing the response of the system to a change in external confinement.
This is probably the most straightforward way to identify the interaction-induced Mott insulator and to distinguish it, e.g., from a disorder induced Anderson insulator~\cite{Anderson:1958a,Billy:2008,Roati:2008}. In a solid, however, the corresponding compressibility can usually not be measured directly, since a compression of the crystalline lattice by an external force does not change the number of electrons per unit cell (see also SOM.1). 

In this work, non-interacting and repulsively interacting spin mixtures of fermionic atoms deep in the degenerate regime are studied in a three-dimensional optical lattice. In the experiment, we are able to independently vary the interaction strength between the fermions using a Feshbach resonance, as well as the lattice depth and the external harmonic confinement of the quantum gas. By monitoring the in-trap density distribution of the fermionic atoms for increasing harmonic confinements, we directly probe the compressibility of the many-body system. This measurement allows us to clearly distinguish compressible metallic phases from globally incompressible states and reveals the strong influence of interactions on the density distribution. For non-interacting clouds the system changes continuously from a purely metallic state into a globally incompressible band-insulating state with increasing confinement. For repulsive interactions, we find the cloud size to be significantly larger than in the non-interacting case, indicating the resistance of the system to compression. For strong repulsion, the system evolves from a metallic state into a Mott insulating state and eventually a band insulator as the compression increases. 

In order to probe the local on-site physics of the system, we measure the fraction of atoms on doubly occupied lattice sites for different experimental parameters. For medium compressions and strong repulsive interactions, we find an almost order of magnitude suppression of the fraction of doubly occupied sites compared to the non-interacting case. For strong compressions the systems enters a band-insulating state and the fraction of atoms on doubly occupied sites becomes comparable for all interactions. In previous experiments, a suppression of the number of doubly occupied sites was demonstrated for increasing interaction strength for bosons~\cite{Gerbier:2006a} and fermions~\cite{Joerdens:2008a} at fixed harmonic confinement, signaling the entrance into a strongly interacting regime. 

The experimentally observed density distributions and fractions of doubly occupied sites are compared to numerical calculations using Dynamical Mean Field Theory (DMFT)~ \cite{Georges:1996,Kotliar:2004,Helmes:2008a,DeLeo:2008}. DMFT is a central method of solid state theory and is widely used to obtain ab-initio descriptions of strongly correlated materials~\cite{Kotliar:2004}. This comparison of DMFT predictions with experiments on ultracold fermions in optical lattices constitutes the first parameter-free experimental test of the validity of DMFT in a three-dimensional system.

\paragraph*{Theoretical model.} 

Restricting our discussion to the lowest energy band of a simple cubic 3D optical lattice, the fermionic quantum gas mixture can be modeled via the Hubbard-Hamiltonian~\cite{Hubbard:1963a} together with an additional term describing the potential energy due to the underlying harmonic potential:

\begin{align*}
\hat{H}= &-J\sum_{\left\langle         i,j\right\rangle,\sigma}\hat{c}_{i,\sigma}^\dagger\hat{c}_{j,\sigma}
+U\sum_i\hat{n}_{i,\downarrow} \hat{n}_{i,\uparrow} \notag \\
&+V_t\sum_i
(i_x^2+i_y^2+\gamma^2 i_z^2)\left(\hat{n}_{i,\downarrow}+\hat{n}_{i,\uparrow}\right).
\end{align*}

Here the indices $i,j$ denote different lattice sites in the three-dimensional system $(i=(i_x,i_y,i_z))$, $\langle
i,j\rangle$ neighboring lattice sites, $\sigma\in \{\downarrow,\uparrow\}$ the two different spin states, $J$ the tunneling matrix element and $U$ the effective on-site interaction. The operators $\hat{c}_{i,\sigma} (\hat{c}_{i,\sigma}^\dagger)$ correspond to the annihilation (creation) operators of a fermion in spin state $\sigma$ on the $i$th lattice site and $\hat{n}_{i,\sigma}$ counts the number of corresponding atoms on the $i$th lattice site.  The strength of the harmonic confinement is parameterized by the energy offset between two adjacent lattice sites at the trap center $V_t=\frac{1}{2}m\omega_\bot^2 d^2$, with $\omega_\perp=\omega_x=\omega_y\neq\omega_z$ being the horizontal trap frequency and $d$ the lattice constant. 
The constant aspect ratio of the trap is denoted by $\gamma=\omega_z/ \omega_\bot$.
Due to the Pauli principle every lattice site can be occupied by at most one atom per spin state.

The quantum phases of the Hubbard model with harmonic confinement are governed by the interplay between three energy scales: kinetic energy, whose scale is given by the lattice bandwidth $12J$ in three dimensions, interaction energy $U$, and the strength of the harmonic confinement, which can conveniently be expressed by the \textit{characteristic trap energy} $E_{t}=V_t(\gamma N_\sigma/(4\pi/3))^{2/3}$, which denotes the Fermi-energy of a non-interacting cloud in the zero-tunneling limit with $N_\sigma$ being the number of atoms per spin state ($N_\downarrow=N_\uparrow$).  The characteristic trap energy depends both on atom number and trap frequency via $E_{t}\propto\omega_\perp^2 N_\sigma^{2/3}$ and describes the effective compression of the quantum gas, which is controlled in the experiment
by changing the trapping potential.

\begin{figure}
\begin{center}
\includegraphics[width=0.9\columnwidth]{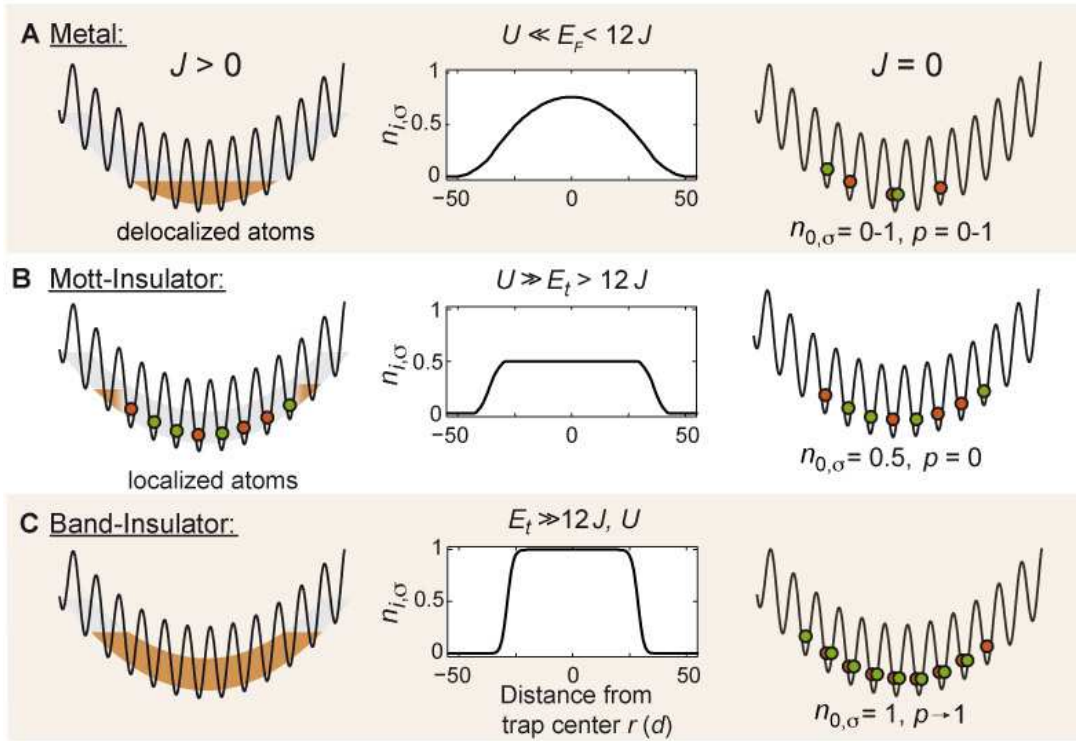}
\caption{Relevant phases of the Hubbard model with an inhomogeneous trapping potential for a spin mixture at $T=0$. A schematic representation is shown in the left column. The center column displays the corresponding in-trap density profiles and the right column outlines the distribution of singly and doubly occupied lattice sites after a rapid projection into the zero tunneling limit $J=0$, with $p$ denoting the total fraction of atoms on doubly occupied lattice sites.  
\label{fig:intro}}
\end{center}
\end{figure}

Depending on which term in the Hamiltonian dominates, different kinds of many-body ground states can occur in the trap center (see Fig.~\ref{fig:intro}). For the case of weak interactions in a shallow trap $U\!\ll\! E_t\!\ll\! 12J$ the Fermi energy is smaller than the lattice bandwidth $(E_F<12J)$ and the atoms are delocalized in order to minimize their kinetic energy. This leads to compressible metallic states with central filling $n_{0,\sigma}\!<\!1$ (Fig.\ref{fig:intro}A), where the local filling factor $n_{i,\sigma}=\langle\hat{n}_{i,\sigma}\rangle$ denotes the average occupation per spin state of a given lattice site. 
A dominating repulsive interaction $U\!\gg\!\!12J$ and $U\!\gg\!E_t$ suppresses the double occupation of lattice sites and can lead to either Fermi-liquid ($n_{0,\sigma}<1/2$) or Mott-insulating ($n_{0,\sigma}=1/2$) states in the center of the trap (Fig.\ref{fig:intro}B), depending on the ratio of kinetic to characteristic trap energy.
Stronger compressions lead to higher filling factors, ultimately ($E_{t}\!\gg\!\!12J$,\,$E_{t}\!\gg\!\!U$) resulting in an incompressible band insulator with unity central filling  at $T=0$  (Fig.\ref{fig:intro}C). 

\begin{figure}
\begin{center}	
\includegraphics[width=1\columnwidth]{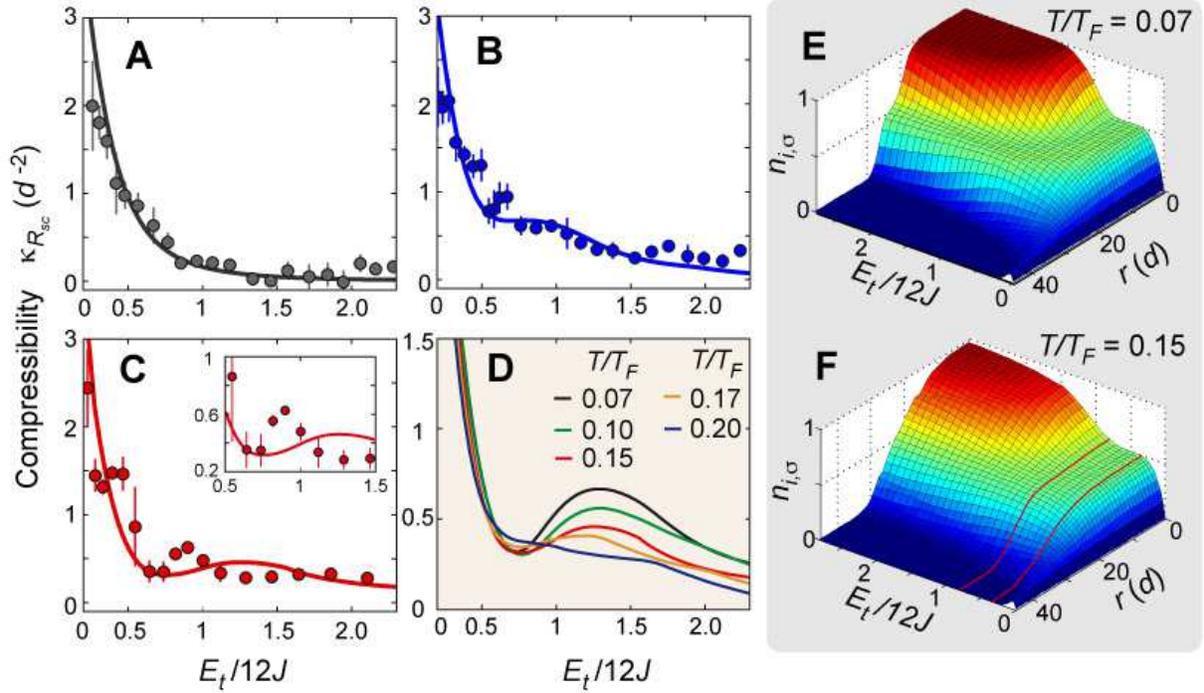}
\caption{
Compressibility and in-trap density distribution.
(\textbf{A-C}) Global compressibility $\kappa_{R_{sc}}$ of the atom cloud for various interactions. Dots denote the result of linear fits on the measured data with the error bars being the fit uncertainty for three interaction strengths ((\textbf{A}) $U/12J=0$, (\textbf{B}) $U/12J=1$, (\textbf{C}) $U/12J= 1.5)$ Solid lines display the theoretically expected results for an initial temperature of $T/T_F=0.15$. The influence of the initial temperature on  the calculated compressibility is shown in \textbf{D} for $U/12J=1.5$. The corresponding density distributions are plotted in \textbf{E, F} with $r$ denoting the distance to the trap center (see also SOM.8). The red lines mark the region where a Mott-insulating core has formed in the center of the trap and the global compressibility is reduced.
\label{fig:comp}}
\end{center}
\end{figure}

Finite temperature reduces all filling factors and enlarges the cloud size, as the system needs to accommodate the corresponding entropy. Furthermore, in the trap the filling always varies smoothly from a maximum at the trap center to zero at the edges of the cloud. For a dominating trap and strong repulsive interaction at low temperature ($E_{t}\!>\!U\!>\!12J$), the interplay between the different terms in the Hamiltonian gives rise to a wedding-cake like structure (see also Fig.~\ref{fig:comp}E, F) consisting of a band-insulating core
($n_{0,\sigma}\approx1$) surrounded by a metallic shell ($1/2\!<\!n_{i,\sigma}\!<\!1$), a Mott-insulating shell ($n_{i,\sigma}=1/2$) and a further metallic shell ($n_{i,\sigma}\!<\!1/2$)~\cite{Helmes:2008a}. The outermost shell remains always metallic, independent of interaction and confinement, only its thickness varies (see also SOM.8).

\paragraph*{Experimental setup.}
Our experiments use an equal mixture of quantum degenerate fermionic $^{40}$K atoms in the two hyperfine states $|F,m_F \rangle=|\frac{9}{2},-\frac{9}{2} \rangle \equiv \spindown$ and $|\frac{9}{2},-\frac{7}{2}\rangle \equiv \spinup$.  
By overlapping two horizontal, elliptical laser beams ($\lambda=1030$\,nm) a pancake-shaped dipole trap with an aspect ratio $\gamma\approx4$ is formed. Using evaporative cooling in this trap, we reach  temperatures of $T/T_{F}=0.15(3)$ with $1.5-2.5\times 10^5$ potassium atoms. A Feshbach resonance located at 202.1\,G~\cite{Regal:2003a} is used to tune the scattering length between the two spin states and thereby control the on-site interaction $U$. The creation of the spin mixture and the last evaporation step are performed either above the resonance (220\,G), giving access to non-interacting (209.9\,G) as well as repulsively interacting clouds with $a\leq 150\,a_0$ ($B\leq 260$\,G), or below the resonance (165\,G), where larger scattering lengths up to $a=300\,a_{0}$ (191.3\,G) can be reached. We have found that a further approach to the Feshbach resonance is hindered by enhanced losses and heating in the lattice (see~\cite{foot3}).

After evaporation, the dipole trap depth is again slightly increased and the magnetic field is tuned to the desired value. Subsequently, a blue detuned 3D optical lattice ($\lambda_{lat}=738$\,nm) with simple cubic symmetry is linearly increased within 10\,ms to a potential depth of $V_{lat}=1\,E_r$. Here $E_r=h^2/(2 m \lambda_{lat}^2)$ denotes the recoil energy, which sets a natural energy scale for the depth of the optical lattice~\cite{Bloch:2008c} and $m$ is the mass of a single atom. The combination of a red-detuned optical dipole trap and a blue-detuned lattice potential allows us to vary lattice depth and external confinement independently. Thereby, the compression of the atomic cloud can be adjusted over a wide parameter range and horizontal trap frequencies as low as $\omega_\perp \simeq 2\pi \times 20$\,Hz can be reached in the presence of the lattice, thus allowing for metallic states with high atom numbers. To monitor the in-situ density distribution under different external confinements, we then ramp the dipole trap depth in 100\,ms to the desired external harmonic confinement ($\omega_{\perp}=2\pi \times 20-120$\,Hz), followed by a linear increase of the optical lattice depth to $V_{lat}=8\,E_r$ in 50\,ms.

An in-situ image of the cloud is subsequently taken along the short axis of the trap using phase-contrast imaging~\cite{Andrews:1996a} at detunings of $\Delta=2\pi\times(200-330$)\,MHz after a hold time of 12\,ms in the lattice. From this picture the cloud size $R=\sqrt{\left\langle r_\bot^2\right\rangle}$ is extracted using adapted 2D Fermi fits (see Supporting Online Material (SOM.5)). 

As phase-contrast imaging is non-destructive, it allows us to also measure the quasi-mo\-mentum distribution of the atoms in the same experimental run using a band-mapping technique~\cite{Kastberg:1995,Greiner:2001b,Kohl:2005a}. To this end, the lattice is ramped down in 200\,$\mu$s and a standard absorption image is taken after 10\,ms time-of-flight.

All experimental data are compared to numerical calculations, in which the DMFT equations of the homogeneous model are solved for a wide range of temperatures and chemical potentials using a numerical renormalization group approach~\cite{Bulla:2001,Bulla:2008} (see SOM.7 for details). As shown in~\cite{Helmes:2008b,Helmes:2008a}, the trapped system can be approximated to very high accuracy by the uniform system through a local density approximation (LDA) even close to the boundary between metal and insulator. For a comparison with the experimental results it is convenient to express the cloud size $R$ in rescaled units $R_{sc}=R/(\gamma N_\sigma)^{1/3}$, along with the dimensionless compression $E_t/12J$. In these units, the cloud size depends only on the interaction strength $U/12J$ and the entropy.
In all calculations, we use the entropy determined from a non-interacting Fermi gas in an harmonic trap at an initial temperature $T/T_F$ and assume adiabatic lattice loading

\begin{figure}
\begin{center}
\includegraphics[width=0.8\columnwidth]{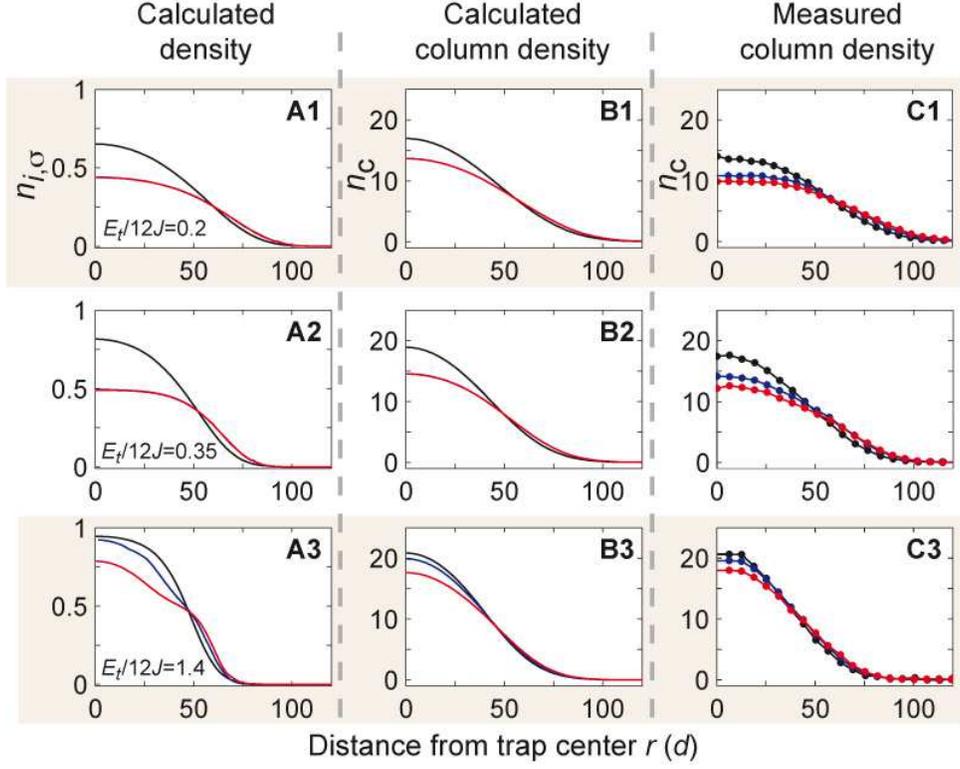}
\caption{Comparison of in-trap density profiles. Calculated radial density profiles for different compressions (harmonic confinements) $E_t/12J$ (left column {\bf (A1-A3)}), corresponding column densities obtained after integration over the $z-$axis and convolution with the point spread function of our imaging system (center column {\bf (B1-B3)}) and experimental results (azimuthally averaged over $>5$ shots) 
for three different interaction strengths $U/12J=0$ (black), $U/12J=1$ (blue) and $U/12J=1.5$ (red) {\bf (C1-C3)}.
At small compressions (A, B) the calculated density profiles for $U/12J=1$ and $U/12J=1.5$ are indistinguishable, as in both cases double occupations are almost completely suppressed for $n_{i,\sigma}<1/2$.
\label{fig:density_prof}}
\end{center}
\end{figure}

A comparison between the numerically calculated density distributions, the corresponding column densities and the experimentally measured ones is presented in Figure~\ref{fig:density_prof} for different confinements and interactions at constant atom number.
At low compression (first row) all shown distributions are purely metallic, as the central filling of $n_{0,\sigma}=0.65$ in the non-interacting case (black curve) is reduced by repulsive interactions to below half filling (blue, red). Due to the fixed atom number, this reduction in density causes an enlargement of the cloud, which is clearly visible in the experiment. In the second row the compression is slightly higher ($E_t/12J=0.35$) and a Mott-insulating core with half filling starts to form for repulsive interactions (A2), while the non-interacting distribution remains metallic. 
For high compressions ($E_t/12J=1.4$, last row) a band-insulating core has formed in the non-interacting case (A3), while the repulsively interacting clouds show a metallic core. For strong interactions (red) the remains of a Mott-insulating ring can be seen (A3). All experimental profiles agree well with theory, small deviations at low compressions could be attributed to an anharmonicity of the confining potential for very shallow traps.

\begin{figure}
\begin{center}
\includegraphics[width=0.8\columnwidth]{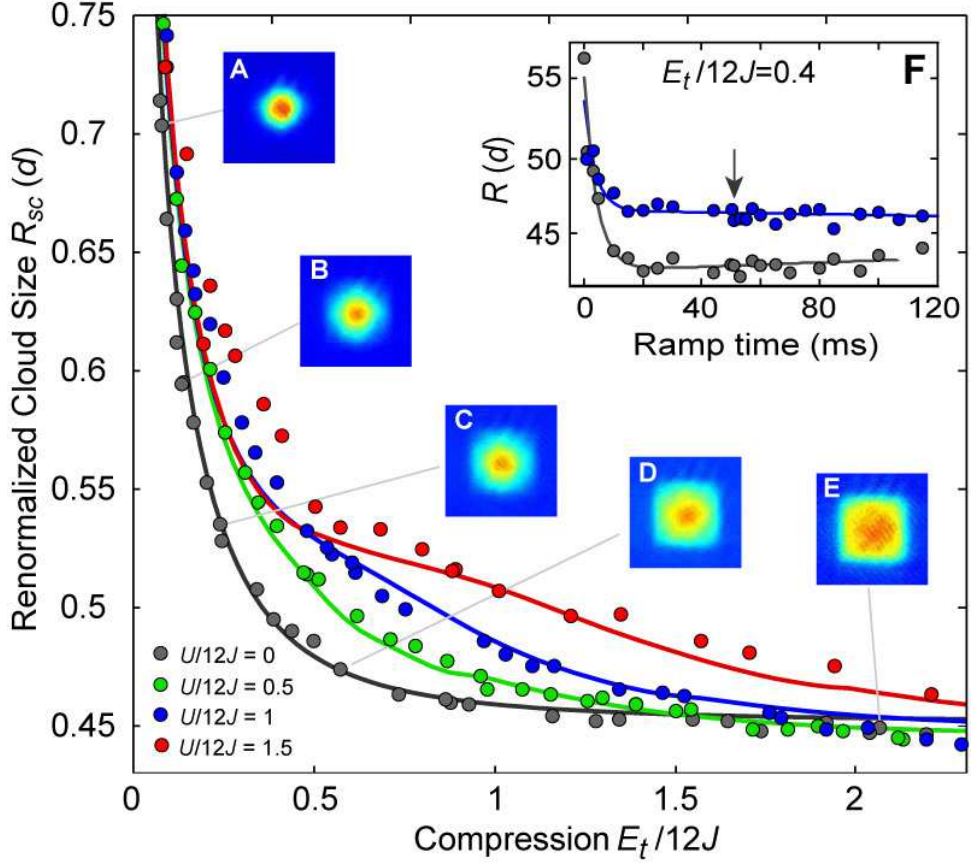}
\caption{Cloud sizes of the interacting spin mixture versus compression. Measured cloud size $R_{sc}$ in a $V_{lat} = 8\,E_r$ deep lattice as a function of the external trapping potential for various interactions $U/12J=0$ (black), $U/12J=0.5$ (green), $U/12J=1$ (blue), $U/12J=1.5$ (red)). Dots denote single experimental shots, lines the theoretical expectation from DMFT for $T/T_F=0.15$ prior to loading. The insets {\bf (A-E)} show the quasi-momentum distribution of the non-interacting clouds (averaged over several shots). {\bf (F)} Resulting cloud size for different lattice ramp times at $E_t/12J=0.4$ for a non-interacting and an interacting Fermi gas. The arrow marks the ramp time of 50\,ms used in the experiment.\label{fig:size}}
\end{center}
\end{figure}

This general behavior can be quantified further by plotting the cloud size $R_{sc}$ (dots) in rescaled units as a function of the characteristic trap energy $E_t\propto\omega_\perp^2 N_\sigma^{2/3}$ (see Fig.~\ref{fig:size}). Additionally, the global compressibility  $\kappa_{R_{sc}}=-\frac{1}{R_{sc}^3} 
\frac{\partial R_{sc}}{\partial(E_t/12J)}$ (see Fig.~\ref{fig:comp} and SOM.1) of the system can be extracted from these measurements using linear fits to four consecutive data points to determine the derivative. 
In the non-interacting case we find the cloud sizes (Fig.~\ref{fig:size}, black dots) to decrease continuously with compression until the characteristic trap energy roughly equals the lattice bandwidth ($E_t/12J \sim 1$). For stronger confinements the compressibility approaches zero (Fig.~\ref{fig:comp}A), as almost all atoms are in the band insulating regime while the surrounding metallic shell becomes negligible. The corresponding quasi-momentum distribution (Fig.~\ref{fig:size}A-E) changes gradually from a partially filled first Brillouin zone, characteristic for a metal, to an almost evenly filled first Brillouin zone for increasing compressions, as expected for a band insulator. We would like to note, however, that a band-mapping technique reveals only the relative occupations of the extended Bloch states. For an inhomogeneous system it therefore yields no information about the real-space density and especially cannot distinguish insulating from compressible states, e.g. non-equilibrium states in which the atomic wavefunctions are localized to single lattice sites. In contrast, the measurements shown here directly demonstrate the global incompressibility of the fermionic band insulator, in excellent agreement with the theoretical expectation for a non-interacting Fermi gas (black line). 

The green, blue  and red dots in figures~\ref{fig:size} and \ref{fig:comp} represent the size and compressibility of repulsively interacting clouds with $U/12J=0.5, 1$ and 1.5 in comparison with the DMFT calculations (lines). 
For moderately repulsive interaction ($U/12J=0.5,1$) the cloud size is clearly bigger than in the non-interacting case but eventually reaches the size of the band insulator. 
For stronger repulsive interactions $(U/12J=1.5)$ we find the onset of a region ($0.5\!<\!E_t/12J\!<\!0.7$) where the cloud size decreases only slightly with increasing harmonic confinement, denoting a very small compressibility, whereas for stronger confinements the compressibility increases again. 
This is consistent with the formation of an incompressible Mott-insulating core with half filling in the center of the trap, surrounded by a compressible metallic shell, as can be seen in the corresponding in-trap density profiles (see  Fig.~\ref{fig:comp}E, F). For higher confinements an additional metallic core ($1/2\!<\!n_{i,\sigma}\!<\!1$) starts to form in the center of the trap. A local minimum in the global compressibility is in fact a genuine characteristic of a Mott-insulator and for large $U$ and low temperature, we expect the global compressibility in the middle of the Mott region to vanish as $1/U^2$ (see SOM.1). The experimental data, indeed, show an indication of this behavior (see Fig.~\ref{fig:comp} C) for increasing interactions. For $E_t/12J\simeq0.5$ a minimum in
the compressibility is observed, followed by an increase of the
compressibility around $E_t/12J\simeq0.8$, slightly earlier than predicted by theory.
 
When the system is compressed even further all cloud sizes approach that of a band insulating state and all compressibilities tend to zero. 
As can be seen in the theory predictions, for strong confinement the repulsively interacting  clouds can, however, become slightly smaller than the non-interacting one due to Pomeranchuk cooling~\cite{Werner:2005}.
At the same average entropy per particle, the interacting system has a considerably lower temperature in the lattice, as the spin entropy is enhanced due to interactions (see SOM.3). 
In the experiment this feature is barely visible, as a second effect becomes important: At very high compressions ($E_t/12J\gtrsim2$) the second Bloch band gets slightly populated
during the lattice ramp up, which leads to smaller cloud sizes for all interactions, as a small number of atoms in a nearly empty band can carry a considerable amount of entropy.

Overall, we find the measured cloud sizes to be in very good quantitative agreement with the theoretical calculations up to $U/12J=1.5$ ($B=175$G). Nevertheless, for repulsive interactions and medium compression ($E_t/12J\approx0.5$) the cloud size is slightly bigger than the theoretical expectation. The discrepancies become more prominent for stronger interactions, i.e.~when tuning the scattering length to even more positive values below the Feshbach resonance. This could be caused by non-equilibrium dynamics in the formation of a Mott-insulating state for strong interactions~\cite{DemlerPrivateComm} or may be an effect not covered by the simple single band Hubbard model or the DMFT calculations and requires further investigation. 

In order to ensure that the used lattice loading time of 50\,ms is adiabatic, we have measured the resulting in-situ cloud size as a function of ramp time (see inset Fig.~\ref{fig:size}F) in the regime around $E_t/12J\approx0.5$, where the differences between experiment and theory are biggest. In this regime the cloud size shrinks during the loading of the lattice, a too fast loading thereby results in a larger cloud. We find the cloud size to decrease significantly for ramp times up to 20\,ms after which no further decrease is observed, which indicates adiabaticity for the used ramp time. However, a second longer timescale, which could become more relevant for stronger interactions~\cite{Winkler:2006}, cannot be ruled out. In addition, the temperatures before loading into the lattice and after a return to the dipole trap with a reversed sequence are compared. We find a rise in temperature between $0.010(5) \,T/T_{F}$ for a non-interacting cloud and $0.05(2)\,T/T_{F}$ for a medium repulsion of $U/12J=1$ at compressions around $E_t/12J\approx0.5$. Even for the highest compressions the temperature increase stays below $0.11(2)\,T/T_{F}$. 
Taking these temperature increases into account, the good agreement between
the experimental data and the numerical calculations, which assume adiabatic loading and an initial temperature of
$T/T_F=0.15$, indicates that our actual initial temperatures lie rather at the lower end 
of the measured temperature range $T/T_F=0.15(3)$.

The theoretical calculations for the compressibility shown in Fig.~\ref{fig:comp}D demonstrate that the entrance into a Mott insulating state, signaled by a minimum in the global compressibility, happens rather rapidly for initial temperatures in the range of $0.15 \lesssim T/T_F \lesssim 0.2$. At these temperatures, the entropy per particle is larger than $k_B\,2\ln2$, which is even larger than the maximum possible entropy per particle of a half-filled Hubbard model in the homogeneous system (reached only at temperatures in the lattice $k_B T_{lat}\gg U$).
In the trap, however, a large fraction of the entropy is accumulated in the metallic shells at the edges of the atomic cloud where the diluted atoms have a large configurational entropy (see SOM.4). Therefore the temperature remains on the order of $k_B T_{lat}\simeq J \ll U$ in the Mott insulating regime (see SOM.3). 

\begin{figure}
\begin{center}
\includegraphics[width=0.7\columnwidth]{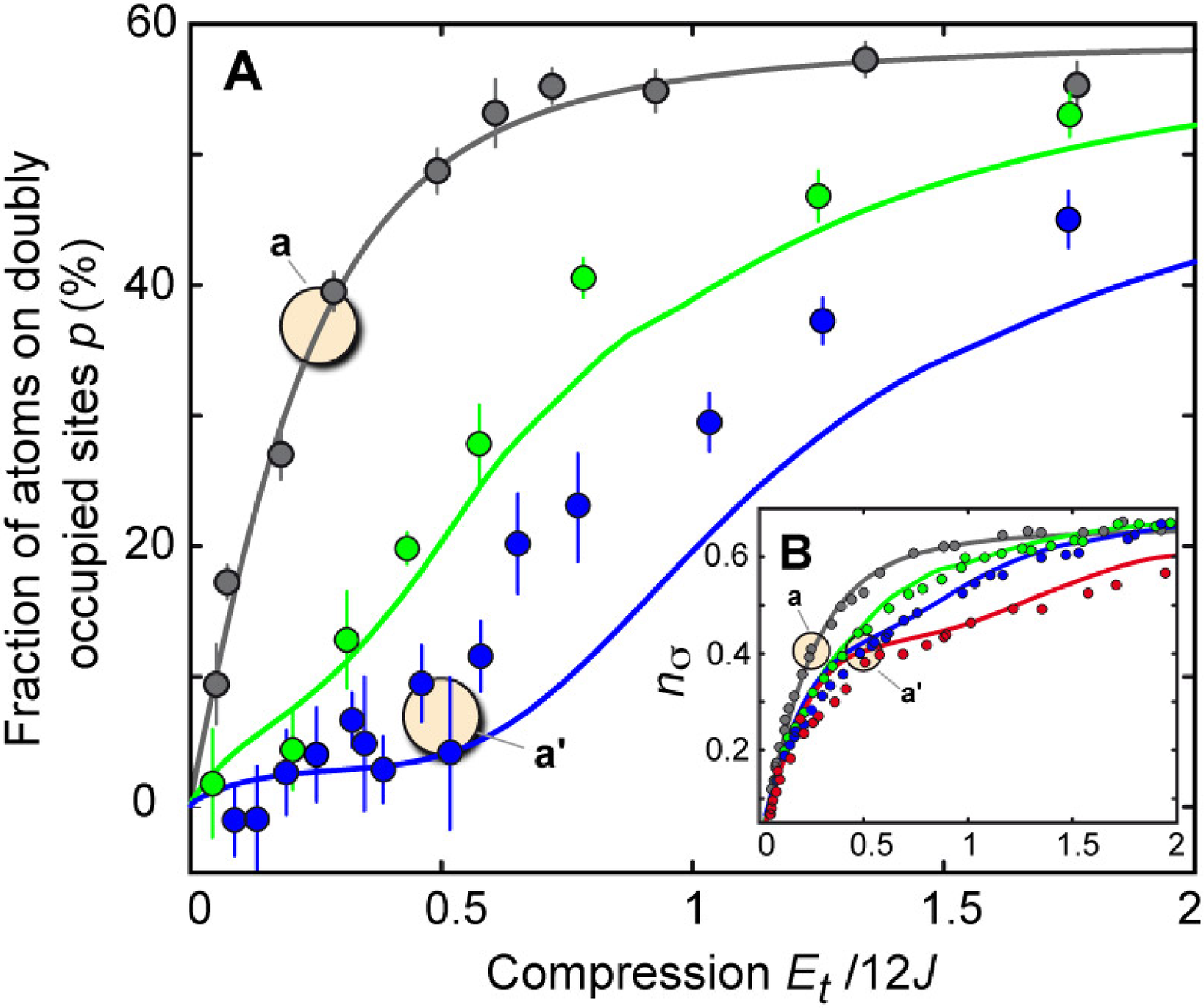}
\caption{Fraction of atoms on doubly occupied sites versus compression for different interaction strengths (black $U/12J=0$, green $U/12J=0.5$, blue $U/12J=1$) {\bf (A)}. The inset {\bf (B)} displays the evolution of the average density for increasing compression and different interaction strengths (as in {\bf A}, red $U/12J=1.5$), allowing one to compare the fraction of doubly occupied sites at constant average density. The yellow circles denoted by {\bf a} and {\bf a'} indicate the fraction of atoms in doubly occupied sites for a constant average density of $n_\sigma=0.4$ and $U/12J=0$ and $U/12J=1$. \label{fig:Molfrac}}
\end{center}
\end{figure}

In addition to the global compressibility measurements, the fraction of atoms on doubly occupied lattice sites (pair fraction $p$) has been measured for magnetic fields above the Feshbach resonance ($U/12J=0,0.5,1$) by converting all atoms on doubly occupied sites into molecules using a magnetic field ramp (0.2\,ms/G) over the Feshbach resonance~\cite{Regal:2003a,Joerdens:2008a}. In this case, the lattice depth is further increased to $20\,E_r$ in $200\,\mu$s to prevent tunneling of the atoms during the field ramp. After the ramp, the lattice depth is linearly decreased to zero in $200\,\mu$s and the number of remaining atoms is recorded by time-of-flight absorption imaging, yielding the number of singly occupied lattice sites. The difference in atom number with and without the magnetic field ramp normalized to the total atom number gives the desired fraction $p$, which is plotted in Fig.~\ref{fig:Molfrac}A including corrections which account for atom losses during the measurement sequence (see also SOM.6).
The fraction of atoms on doubly occupied sites gives insight into the local on-site physics of the system. 
In combination with the in-situ size measurements, this fraction can be compared for different interaction strengths at constant average density of the system. The average density  $n_\sigma=(\gamma\,N_\sigma/\frac{4}{3}\pi R^3)\times1/n_{BI}$, where $n_{BI}$ denotes the average density of a pure band-insulator at $T=0$, can be calculated from the recorded cloud size $R$ and is shown in Fig.~\ref{fig:Molfrac}B.

In the limit of very low compression (weak confinement) the average density is small $n_\sigma<0.1$ and $p$ tends to zero.  For intermediate densities the fraction of doubly occupied sites depends crucially on the interaction. At a constant mean density of $n_\sigma=0.4$, we find a pair fraction of 40\% for a non-interacting cloud and around 5\%  for a medium repulsive interaction $U/12J=1$ (see yellow circles in Fig.~\ref{fig:Molfrac}A,~B). In this regime of repulsive interactions at medium compression it is energetically favorable to reduce the number of doubly occupied sites although this costs potential and kinetic energy. As a consequence, different compressions are needed to reach the same average density for different interactions, as can be seen in Fig.~\ref{fig:Molfrac}B.

For strong compressions, the measured pair fraction becomes comparable for all interactions (see Fig.~\ref{fig:Molfrac}), as the potential energy in the trap becomes larger compared to $U$ 
and all atom distributions are expected to contain a large band-insulating core (see SOM.8).
Ultimately, the pair fraction (average density) is limited to values smaller than 60\% ($n_\sigma<0.7$) due to the finite entropy per particle, which reduces the filling factor in the band-insulating state.  While the non-interacting and slightly repulsively interacting curve $U/12J=0.5$ match the DMFT results for an inital temperature of $T/T_{F}=0.15$, we see deviations for stronger repulsive interactions ($U/12J=1$). In this case the measured pair fraction is in general $\simeq10\%$ higher than predicted by theory, nevertheless the qualitative behavior agrees very well. We note that a suppressed pair fraction in comparison with the non-interacting case occurs for all temperatures in the lattice below $k_B\,T_{lat}\approx U$, regardless of the formation of an incompressible Mott insulating phase in the inhomogeneous system (see SOM.2). Furthermore, the pair fraction vanishes even for a compressible purely metallic phase with $n_{i,\sigma}<1/2$ in the strongly interacting regime.

\paragraph*{Summary and Outlook.}         
In conclusion, we have investigated a spin mixture of repulsively interacting fermionic atoms in a 3D optical lattice. Using a novel measurement technique, we have been able to directly determine the global compressibility of the many-body quantum system and have explored the different regimes of the interacting mixture from a Fermi liquid to a Mott and band insulating state upon increasing harmonic confinements and increasing interactions. Additionally, the fraction of atoms on doubly occupied sites has been recorded in order to determine the local on-site behavior of the system for a large range of compressions. Our measurements are in good agreement with the results predicted by DMFT for interaction strengths up to $U/12J=1.5$ and initial temperatures of the mixture of $T/T_F=0.15$. For stronger interactions, we find deviations from the expected cloud radius in the low compression regime, which could point to interesting non-equilibrium dynamics in the loading process~\cite{DemlerPrivateComm}, that hinders approaching the expected equilibrium many-body state of the system in this parameter regime. Our numerical calculations also show that low initial temperatures of $T/T_F \lesssim 0.15$ are needed to enter an incompressible many-body Mott insulating state, whereas a suppressed pair fraction compared to the non-interacting state can already be observed at much higher temperatures and for metallic states. This behavior is in fact similar to the results obtained in recent calculations and experiments on the melting of incompressible bosonic Mott insulating shells for increasing temperatures~\cite{Gerbier:2007b,Foelling:2006}.
 
Our measurements are an important step in the direction of analyzing fermionic many-body systems with repulsive interactions in a lattice. For initial entropies lower than $S/N\lesssim k_B\,\ln 2$, one expects the system to enter an antiferromagnetically ordered phase, as the temperature of the quantum gas can then drop below the superexchange coupling that mediates an effective spin-spin interaction between the particles~\cite{DeLeo:2008,Werner:2005,Koetsier:2008,Snoek:2008a}. This would open the path to the investigation of quantum magnetism with ultracold atoms~\cite{Lewenstein:2007}, being an encouraging starting point to ultimately determine the low-temperature phase diagram of the Hubbard model~\cite{Lee:2006,Hofstetter:2002}. This  includes the search for a $d$-wave superconducting phase~\cite{Trebst:2006} that is believed to emerge from within the two-dimensional Hubbard model. 

\clearpage

\clearpage

\clearpage 

\clearpage

\pagebreak
\bibliography{RepulsiveFermions_21}
\bibliographystyle{Science}

% Following is a new environment, {scilastnote}, that's defined in the
% preamble and that allows authors to add a reference at the end of the
% list that's not signaled in the text; such references are used in
% *Science* for acknowledgments of funding, help, etc.

\begin{scilastnote}
\item We acknowledge help during the setup of the experiment by D.~van~Oosten, T.~Rom and S.~Braun and thank the Leibniz Institute of Surface Modification (IOM Leipzig) for the production of phase masks used in phase-contrast imaging. Furthermore, we acknowledge funding by the DFG (FOR801, SFB608, SFB TR 12), the EU (QUASICOMBS, SCALA), AFOSR, DARPA (OLE) and supercomputer support by the John von Neumann Institute for Computing (J\"ulich). S.W. acknowledges additional support by MATCOR. 
\end{scilastnote}

\pagebreak

\begin{center}\huge Metallic and Insulating Phases of Repulsively Interacting Fermions in a 3D Optical Lattice
\end{center}

\begin{center}
\huge Supporting online materials 
\end{center}

\section{Compressibility of atom clouds}

The concept of compressibility is used in rather different ways in
physics. Often the term refers to the change of volume when
pressure is applied, $\kappa_V=-\frac{1}{V} \frac{\partial V}{\partial
  p}$.  To investigate the effects of strong interactions on various phases, a
different concept, usually called electronic compressibility, is very
useful.  Here one tracks the change of the number of electrons per
unit cell $n$ of a solid as a function of the chemical potential,
$\kappa_n=\frac{1}{n^2}\frac{\partial n}{\partial \mu}$. A central
property of a Mott insulator is its incompressibility: $\kappa_n$
vanishes exponentially for low temperatures. The compressibility allows one
to distinguish a Mott insulator, e.g., from an Anderson insulator of
non-interacting particles in the presence of strong
disorder. $\kappa_n$ can be measured only indirectly in
three-dimensional solids, e.g. by comparing shifts in photoemission
spectra~{\it (S1)} for samples with different
doping. The interpretation of such an experiment is, however, difficult, as
a doping of a sample also changes other contributions to the total
energy (e.g. the Coulomb energy) which also lead to a shift of the spectra. A more direct
measurement of $\kappa_n$ can be obtained in thin films or field
effect transistors where the
density of electrons can be changed using a back gate, see e.g.~{\it (S2)}.

It is important to realize that in solids the compressibility of the
volume $\kappa_V$ is usually unrelated to the electronic
compressibility $\kappa_n$ as the number of electrons per unit cell is
not modified when pressure is applied. The situation is completely
different for atomic clouds. When an atomic cloud is compressed
by a change of the trapping potential $V_t$ the density
of atoms does change considerably as the radius of the cloud $R$ will be
reduced. Therefore the compression of an atom cloud gives valuable
information on the properties of the various phases realized in the trap.

The global compressibility of the cloud can be defined as $\kappa_R=-\frac{1}{R^3} 
\frac{\partial R}{\partial V_t}$, where  
$R=\sqrt{\int r^2 n_c(r) d^2 r/N}$ is the typical radius of the cloud
($N=2 N_\sigma$) and $n_c(r)$ is the column density. 
For simplicity, we use an isotropic trap ($\gamma=1$) in the following. 
The powers of $R$ in the
prefactors are chosen such that the quantities are well-defined in the
thermodynamic limit
$N_\sigma \to \infty$ for fixed $E_t$. In rescaled units the corresponding definition
is $\kappa_{R_{sc}}=-\frac{1}{R_{sc}^3} 
\frac{\partial R_{sc}}{\partial(E_t/12J)}=12J(\frac{4}{3}\pi)^{2/3}\kappa_R$

As has been shown in~{\it (S3,S4)}, 
the local density of the system is well described by the density of a
homogeneous system with the chemical potential $\mu-V_t r^2$ (LDA approximation).
When $V_t$ is changed, both the chemical potential $\mu$ and the
temperature $T$ are modified. Using that both particle number and
entropy remain constant,
one can eliminate $\partial \mu/\partial  V_t$ and 
$\partial T/\partial  V_t$ to express the total compressibility of
the trap in terms of grand-canonical derivatives of the density $n$ and entropy $s$
per lattice site
\begin{eqnarray}\label{kappaR}
\kappa_R=-\frac{1}{R^3} 
\frac{\partial R}{\partial V_t}&=&\frac{1}{3 N R^4} \int (r^2-r_0^2) r^2\frac{\partial
  n}{\partial \mu} d^3 r
\end{eqnarray}
where $r_0$ is a typical radius defined by
\begin{eqnarray}
r_0^2=r_c^2+\frac{(r_c^2-r_s^2)^2 \left(\frac{\partial S}{\partial \mu}\right)^2}{r_c^2
  \left( \frac{\partial N}{\partial \mu} \frac{\partial S}{\partial
      T}-\frac{\partial N}{\partial T} \frac{\partial S}{\partial \mu}
  \right)}, 
\end{eqnarray}
with $r_c^2=\int  r^2  \frac{\partial n}{\partial
    \mu} \, d^3 r/\int  \frac{\partial n}{\partial
    \mu} \, d^3 r$ and $r_s^2=\int  r^2  \frac{\partial s}{\partial
    \mu} \, d^3 r/\int  \frac{\partial s}{\partial
    \mu} \, d^3 r$ and the total entropy $S=\int s d^3r$. For low
  temperature one can neglect the rise of the temperature when
$V_t$ is increased and one obtains $r_0\approx r_c$. Under the
experimental conditions, however, one has to take the
heating effects into account to describe the experiments correctly.

Eq.~\ref{kappaR} can be used to estimate the size of $\kappa_R$ deep
in the Mott insulating phase when the Mott insulating core is
surrounded by a compressible shell of width $\Delta r$. The metallic
shell is approximately located at $r_0$ and its width is inversely 
proportional to the slope of the potential, 
$\Delta r \sim J/(V_0 r_0)$. Due to particle number conservation reflected in
the $r_0^2$ counterterm in Eq.~(\ref{kappaR}), $\kappa_R$ is
proportional to $(\Delta r)^2$, $\kappa_R \sim \frac{1}{N R^4 }
\frac{(\Delta r)^2 r_0^5}{J}$. Using that $V_0 r_0^2\sim U$ and $r_0
\sim R \sim N^{1/3}$ we obtain the expected result that $\kappa_R$
vanishes rapidly for large $U$, $\kappa_R \sim J/E_t^2$ reaching
$J/U^2$ in the center of the Mott plateau, see Figure S1. 
When $V_t$ is further
increased, particles from the outer shell can move to the center of
the trap and the counter-term $r_0^2$ becomes much smaller. In this
regime $\kappa_R$ is {\em linear} in $\Delta r$ and therefore
$\kappa_R \sim 1/U$ until the center of the trap is filled with doubly occupied
sites. 
In the low density limit, $V_t
\to 0$, $\kappa_R$ diverges
proportional to $1/\sqrt{V_t}$  at low temperatures as interactions
 can be neglected and
$R\sim N^{1/3} (E_t/12 J)^{-1/4}$. The above described behavior of
$\kappa_R$ is shown in Figure S1.

\begin{figure}[H]
\begin{center}
\includegraphics[width=0.7 \columnwidth,clip]{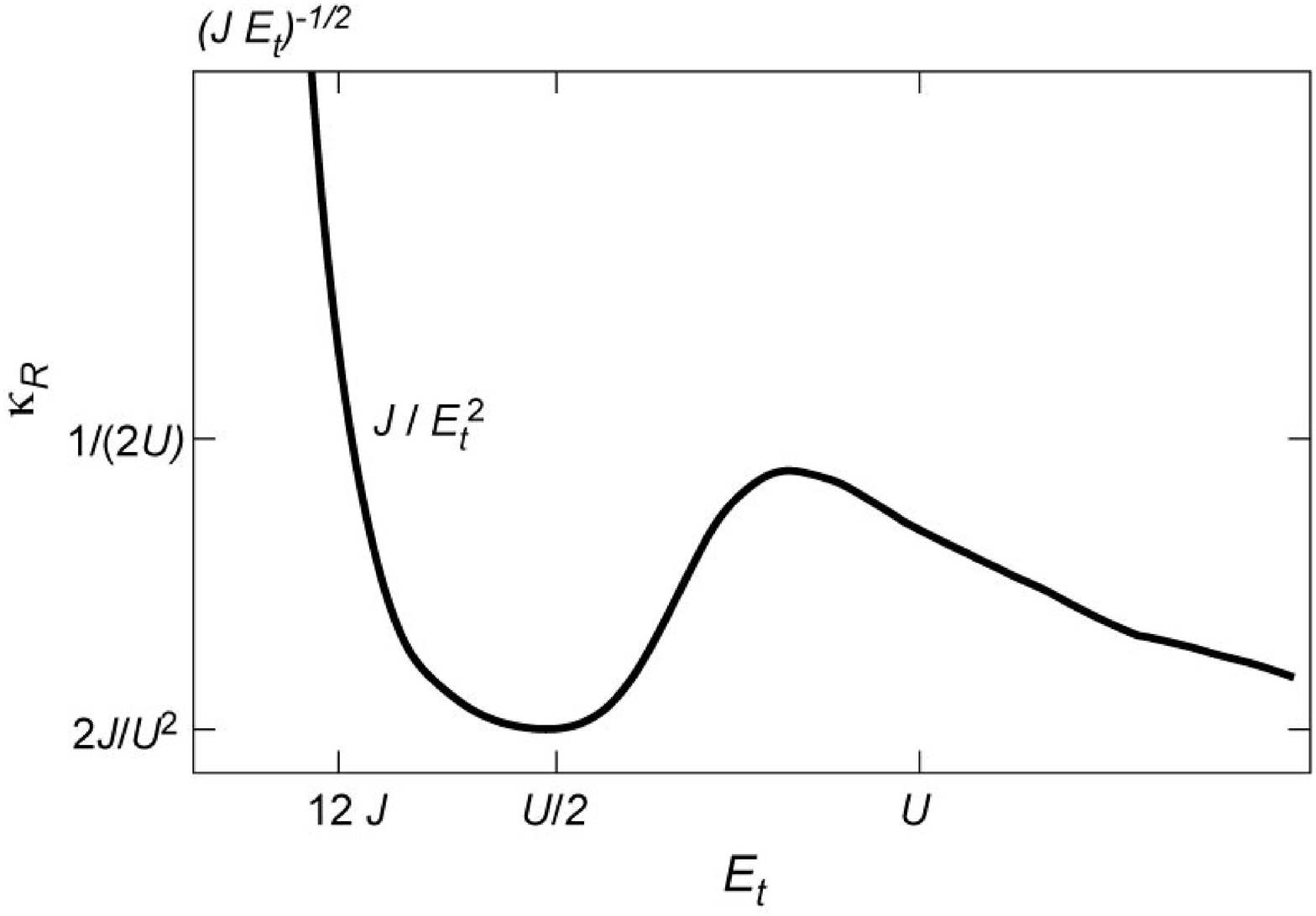}
\caption*{Figure \textbf{S1:} $\kappa_R$ as a function of $E_t$. The qualitative behavior
  is shown using the DMFT results for $U/12 J=2.5$ for an initial
  temperature of $T/T_F=0.15$ as an example. The distinct minimum in the global compressibility $\kappa_R$ is
  a clear signature of the incompressible Mott-insulator. 
  For lower temperatures, the minima and maxima become more pronounced, see Fig.4 of the main paper.}
  \end{center}
\end{figure}
\begin{figure}[H]
\begin{center}
\includegraphics[width=0.6\columnwidth]{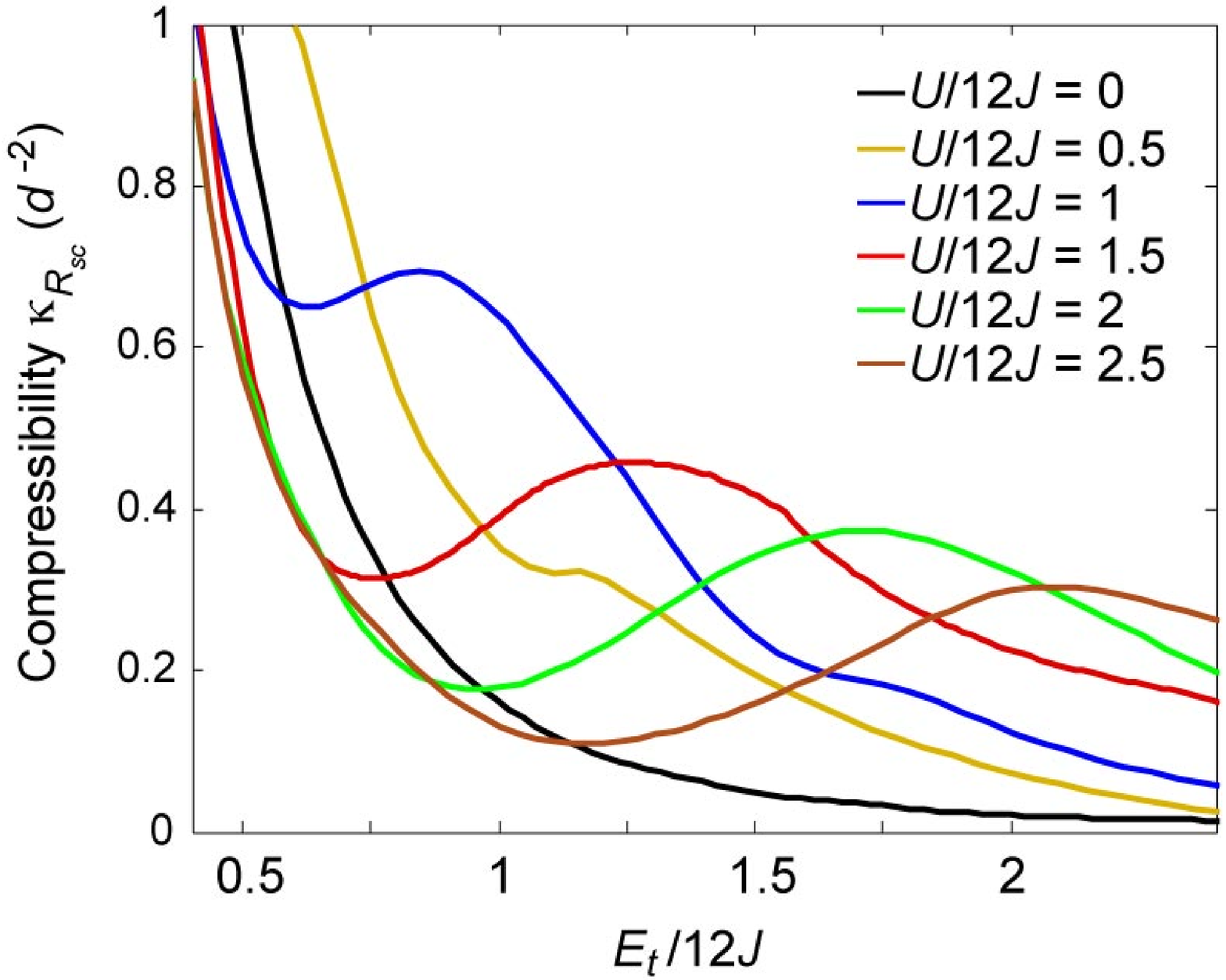}
\caption*{Figure \textbf{S2:} Calculated global compressibility $\kappa_{R_{sc}}$ in rescaled units as a function of $E_t/12J$ for different interaction strengths ($U/12J=0\ldots2.5$). All curves are calculated for the same initial temperature $T/T_F=0.15$ prior to loading of the lattice. 
 }
\end{center}
\end{figure}

Figure S2 shows the calculated global compressibility for various interactions and an initial temperature of $T/T_F=0.15$. Starting at $U/12J=1$ a minimum in the global compressibility appears, as an incompressible Mott-insulator forms in the center of trap (for the uniform system at $T=0$ critical interaction for the metal to insulator transition, within DMFT, is $U/12J = 1.26$~{\it (S4)}). For larger interactions the minimum becomes broader and deeper as the Mott-insulting core increases in size (see density profiles in SOM.8). The remaining finite global compressibility is dominated by the compressible metallic shell around the incompressible Mott-insulator and decreases like $\propto 1/U^2$.

\section{Melting of the Mott-insulator for increasing temperature}
\begin{figure}[H]
\centering
\includegraphics[width=0.5\columnwidth]{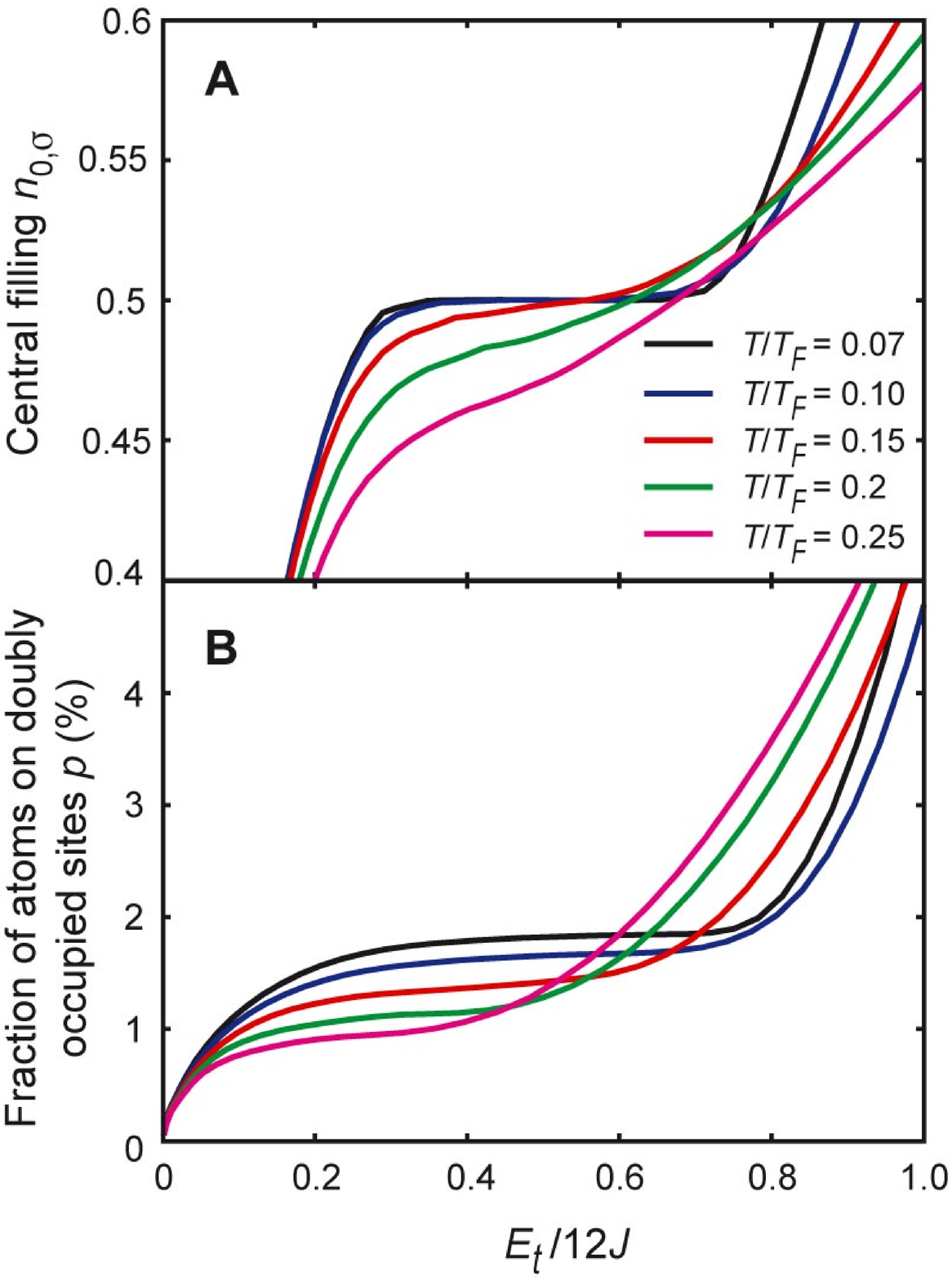}
\caption*{Figure \textbf{S3:} Melting of the Mott insulator with $U/12J=1.5$ for increasing temperature. 
\textbf{A}: calculated central density, \textbf{B}: corresponding pair fraction p.
 }
\end{figure}
In Figure~S3A the central in-trap density is plotted for various initial temperatures as a function of compression.
The vanishing slope at $E_t/12J\approx0.5$ for low enough temperatures directly shows 
the local incompressibility ($\kappa_n\approx0$) of the Mott insulator, 
as the chemical potential increases with higher compressions. For temperatures around $T/T_F\approx0.15$ 
the Mott insulator melts and the local compressibility strongly increases. The pair fraction on the other hand, which is shown in Figure~S3B remains strongly suppressed and changes by less than 1\% over 
the whole temperature range, showing no signature of the crossover into the Mott insulator.

\section{Pomeranchuk cooling}
\begin{figure}[H]
\begin{center}
\subfloat{\includegraphics[width=0.45\columnwidth]{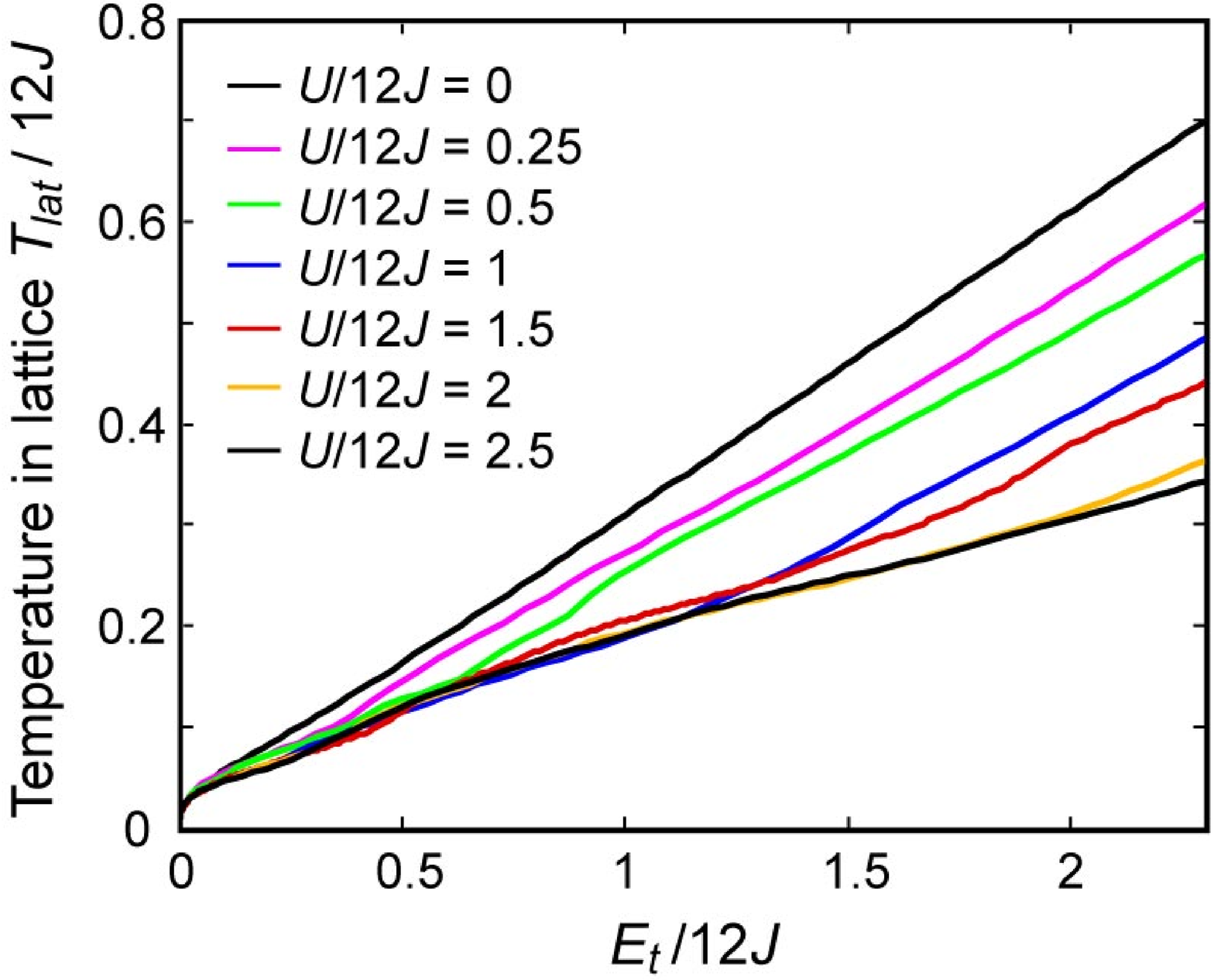}}
\subfloat{\includegraphics[width=0.47\columnwidth]{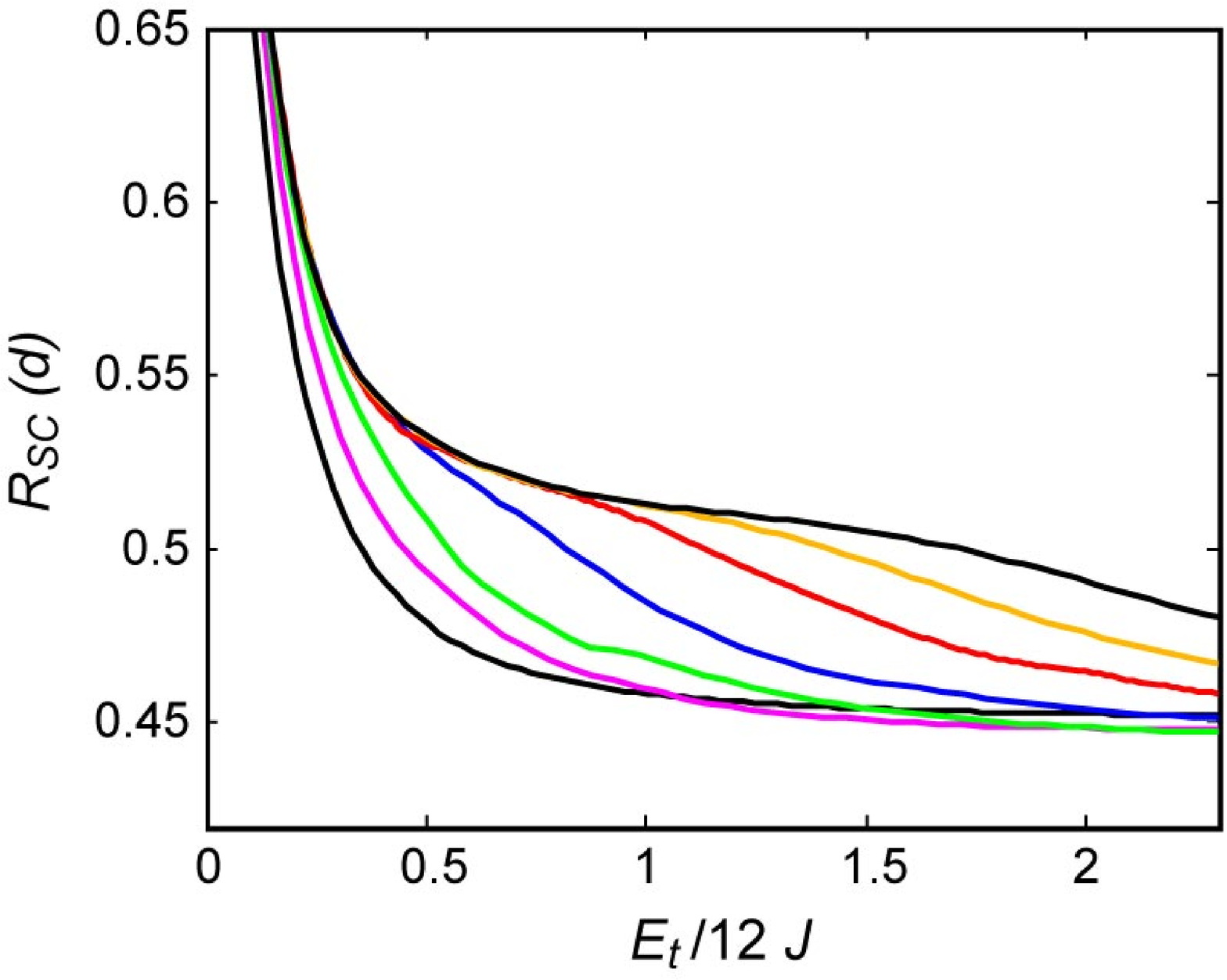}}
\caption*{Figure \textbf{S4:} Calculated temperature and cloud size in the lattice. \textbf{A:} Calculated temperature $T_{lat}/12J$ in the lattice for various interaction strengths at a constant initial temperature of $T/T_F=0.15$ prior to the loading. Small wiggles in the curves are artifacts of the interpolation of the entropy, see SOM.7. \textbf{B:} Calculated cloud size as a function of compression $E_t/12J$ for the same interactions as in \textbf{A}. At compressions larger than one the size of the repulsively interacting clouds becomes smaller than in the noninteracting cloud due to the Pomeranchuk effect.
 }
\end{center}
\end{figure}
The final temperature in the lattice, which is plotted in Figure S4A in rescaled units,  is a function of the initial entropy per particle, the compression and the interaction. Due to the Pomeranchuk effect, the temperature in the lattice $T_{lat}$ decreases for stronger interactions as the particles become more localized, which increases the spin entropy. Therefore the cloud size at high compressions can become slightly smaller for medium interactions than for zero interaction, as can be seen in Figure S4B.

\section{Entropy distribution}
\begin{figure}[H]
\begin{center}
\includegraphics[width=0.7\columnwidth]{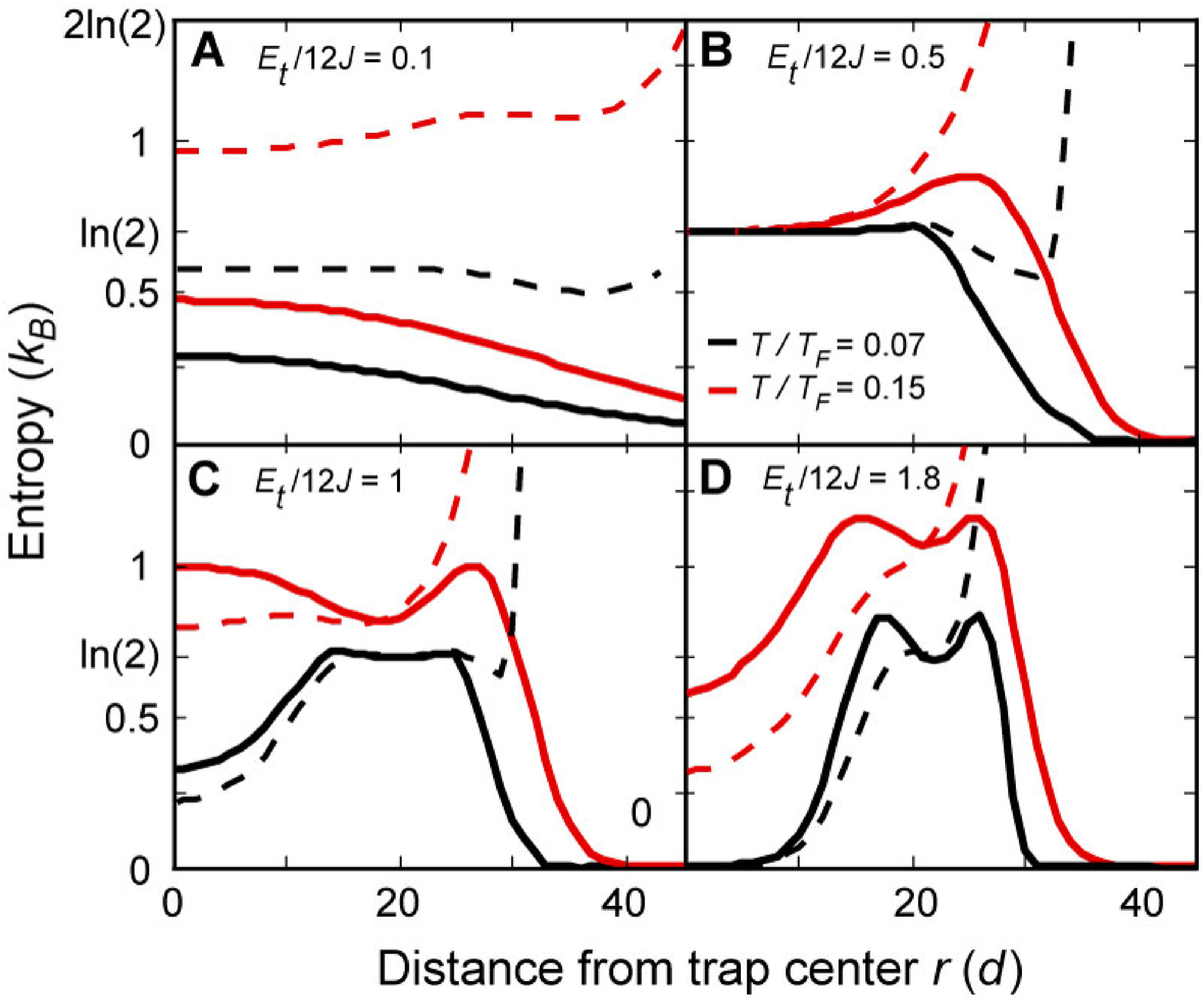}
\caption*{Figure \textbf{S5:} Calculated entropy distributions in the lattice.
Solid (dashed) lines show the entropy per lattice site (per particle) for initial temperatures of  $T/T_F=0.07$ (black) and $T/T_F=0.15$ (red) and strong repulsive interaction $U/12J=1.5$. 
} 
\end{center}
\end{figure}
In Figure S5A the entropy distribution of a purely metallic state with less than half filling in the center of the trap is shown. At higher compression (\textbf{B}) a Mott-insulator with unity filling and $k_B\,\ln2$ entropy per particle has formed in the center of the trap even in the case of $T/T_F=0.15$, for which the average entropy per particle is above $k_B\,2\ln2$. This is possible only due to the inhomogeneity of the system, as most of the entropy is carried by the metallic shells where the entropy per particle can diverge.
For high compressions (\textbf{D}) a band-insulating core has formed and for low enough temperatures (black) nearly all entropy is carried by the surrounding shells. The small dip at $r=20$ is a remnant of the Mott-insulating shell between the two metallic shells. 
Note that entropy per lattice site and entropy per particle are equal at half filling.

\section{Fitting procedure}
To extract the cloud size $R$ from the phase contrast images, we fit the 2D image of the cloud with the following adapted Fermi-Fit function: \[F(x,y)=a\, \text{Li}_2\left(-100\,e^{-\frac{(x-x_c)^2}{2\sigma_x^2}-\frac{(y-y_c)^2}{2\sigma_y^2}}\right) +b+c\,\sqrt{\frac{(x-x_c)^2}{\sigma_x^2}+\frac{(y-y_c)^2}{\sigma_y^2}}\] with $\text{Li}_2$ being the di-logarithm and $x_c,y_c,\sigma_x,\sigma_y,a,b,c$ free fit parameters. The last term models a broad funnel-shaped background, which is an artifact of phase-contrast imaging. As can be seen in Figure S6, this function describes our measured curves much better than a Gaussian. Indeed the adapted Fermi-Fits yield on average 8\% (non-interacting cloud) and 23\% (interacting cloud with $U/12J=1.5$) smaller squared residuals compared to a Gaussian fit function including the last background term. Using this fit-function, the imaged cloud size $R=\sqrt{\left\langle r_\bot^2\right\rangle}$ after integration over the propagation axis of the imaging laser beam is given by $R=\sqrt{{1.264}^{2}\, (\sigma_x^2+\sigma_y^2)-\eta^2}$, where $\eta$ denotes the width of the point spread function of the imaging setup, which is well below one third of the smallest used cloud size.

\begin{figure}[H]
\begin{center}
\includegraphics[width=0.5\columnwidth]{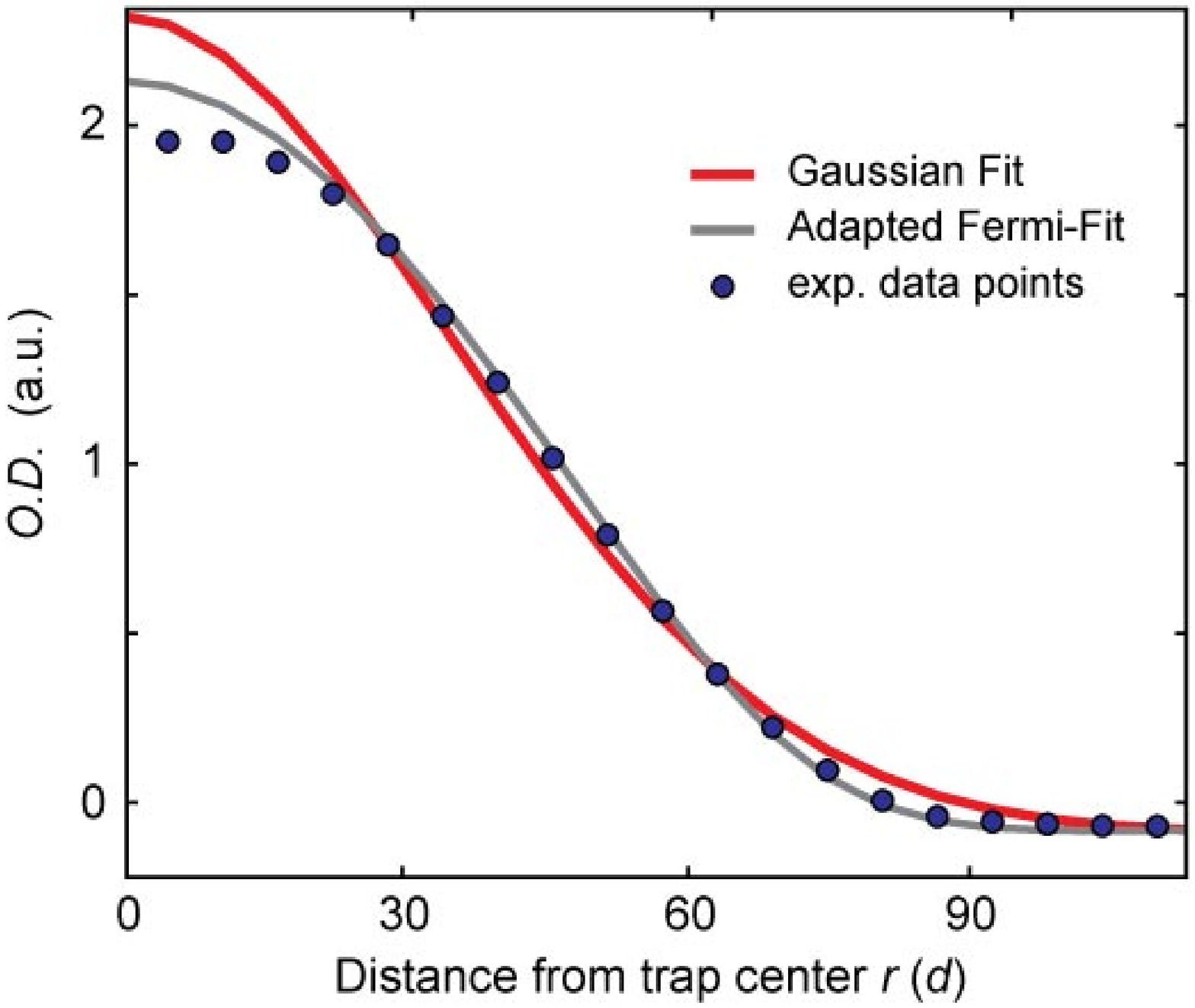}
\caption*{Figure \textbf{S6:} Comparison of different fit functions. The azimuthally averaged data (blue dots) is shown together with a 
fitted Gaussian (red line) and an adapted Fermi-fit (black line). 
Both fits were performed on the full 2D data before averaging.
The influence of the small deviation at the trap center on the resulting atom number (cloud size) is below 1\% (0.1\%), as the statistical weight of each averaged data point increases linearly with the distance $r$ from the cloud center.}
\end{center}
\end{figure}

\section{Atom number calibration and fraction of atoms on doubly occupied sites}
The atom numbers in the manuscript were calibrated such that the theoretically calculated renormalized cloud size for non-interacting atoms at high compressions was reproduced. This atom number calibration agrees within 10\% with both the theoretical expectation when taking the polarization of the probe beam and the corresponding Clebsch-Gordan coefficients of the transition used for phase contrast imaging into account and a second independent calibration, in which the cloud size of a non-interacting Fermi gas in a pure dipole trap was measured as a function of the trap frequency and compared to the theoretically expected scaling.

When measuring the fraction of atoms on doubly occupied lattice sites, we need to take a strong reduction of the pair lifetime for high lattice depths $V_{lat}=20\!-\!25\,E_r$ into account. During the 15\,ms hold time used for the Feshbach ramps, doubly occupied sites can be lost due to a light assisted collision process caused by the blue detuned lattice light~{\it (S5,S6)}. This process selectively affects only doubly occupied sites and therefore yields an underestimation of the measured pair fraction $p$, for which the experimental data has been accordingly corrected. The observed atom losses range from 15\% for non-interacting atoms to below 5\% for strong repulsive interaction and are consistent with the independently measured lifetimes of $70(10)$\,ms ($a=0a_0$) resp. $310(70)$\,ms ($a=150a_0$) for doubly occupied sites in a deep lattice $V_{lat}=24E_r$.

\section{Numerical implementation}
To obtain the theoretical results shown in the paper, we use the
numerical renormalization group method (NRG) to solve the dynamical
mean field equations for the Hubbard model in the thermodynamic limit,
see Refs.~{\it (S7,S8)}.  
The NRG calculations were carried
out using a logarithmic discretization of the conduction band with 
discretization parameter $\Lambda/6J=1.5$, $N_{sh}=78$ 
energy shells and $660$ retained states per energy shell.
From the solution we obtain
the average number of atoms per site and spin, $n(\mu,T)$, and the average number
of doubly occupied sites, $n_d(\mu,T)$, for a dense linear grid of
chemical potentials and a logarithmic grid of temperatures. We
interpolate the result. As the NRG code does not work very well for
extremely low densities of particles or holes, we use the
non-interacting result in this regime for $n(\mu,T)$ (typically
$n<0.03$ for moderate $T$). It is, however, not possible to
approximate $n_d(\mu,T)$ by the non-interacting result in this
regime. For finite temperature, the number of doubly occupied sites is  proportional to
$n^2$
for $n\to 0$ with a proportionality factor which is independent of $T$
in this classical limit. We determine the proportionality factor $c(U)$ numerically
($c(U)\approx 0.39, 0.22, 0.13, 0.087, 0.064$ for values of $U$
ranging from $U/12 J=0.5$ to $2.5$) and use $n_d \approx c(U) n^2$ in
the low-density limit.

A direct calculation of the entropy within
DMFT+NRG is subtle, therefore we use the thermodynamic relation
$\partial n/\partial T=\partial s/\partial \mu$ to obtain
$s(\mu,T)=\int_{-\infty}^\mu  \partial n/\partial T$ for
$\mu<U/2$ and  $s(\mu,T)=s(U-\mu,T)$ for $\mu>U/2$. 
The interpolation and integration errors due to this procedure are
typically less than 5\% as one can check by comparing $s(\mu=U/2)$
with $\ln 2$, the value expected in the Mott insulating phase.
We then use the local density approximation (LDA), justified 
in~{\it (S3,S4)}), to obtain the entropy per particle
$S/N(\mu,T)=\int s(\mu-V_t r^2,T)d^3 r/\int n(\mu-V_t r^2,T)d^3 r$ as a
function of $\mu$ and $T$. Note that $S/N$ does not depend on $V_t$
within LDA. For fixed $V_t,\,\mu$ and $S/N$ we then determine the
temperature $T$ and the particle number $N=\int n(\mu-V_t r^2,T)d^3
r$. Up to a trivial rescaling, all physical
properties depend (within LDA) only on $E_t/(12 J)\propto V_t
N^{2/3}$. We can therefore directly obtain quantities like 
the fraction of atoms on doubly occupied sites $p=N_d/N =\int n_d(\mu-V_t r^2,T)d^3
r/N$, the rescaled column density $n_c(r)/N^{1/3}=\int
n(\mu-V_t(r^2+z^2),T) dz/N^{1/3}$ 
or the effective radius of the cloud $R/N^{1/3}=\sqrt{\int r^2 n_c(r) d^2r/
  \int n_c(r) d^2 r}$ as a function of $E_t/12J$ from the above described procedure. 

\pagebreak
\section{Density distributions}
\enlargethispage{4\baselineskip}\thispagestyle{empty}
\begin{figure}[H]
\begin{center}
\includegraphics[width=1\columnwidth]{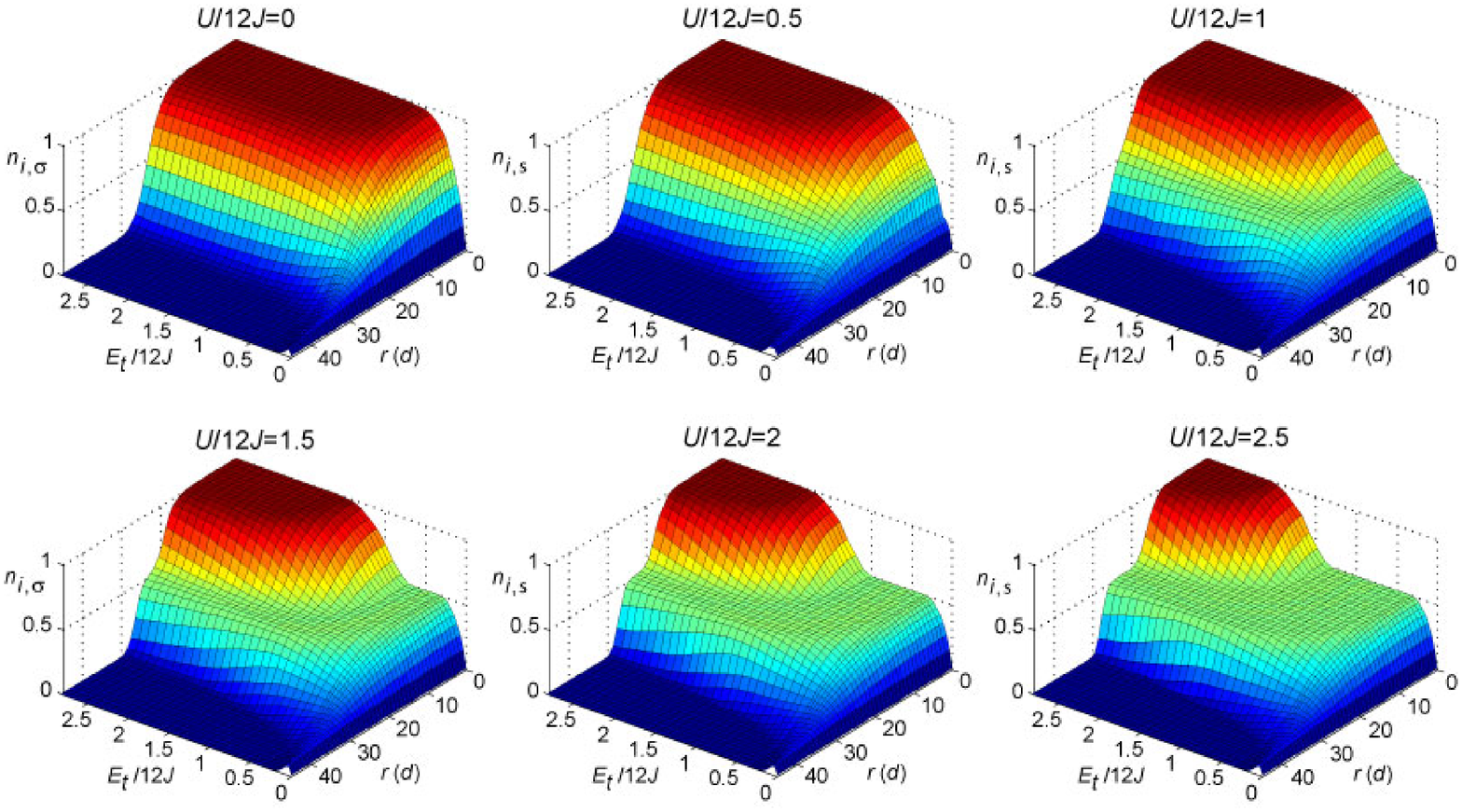}
\caption*{Figure \textbf{S7:} Calculated density distributions for different interactions and an initial temperature of $T/T_F=0.07$.}
\vspace{10pt}
\includegraphics[width=1\columnwidth]{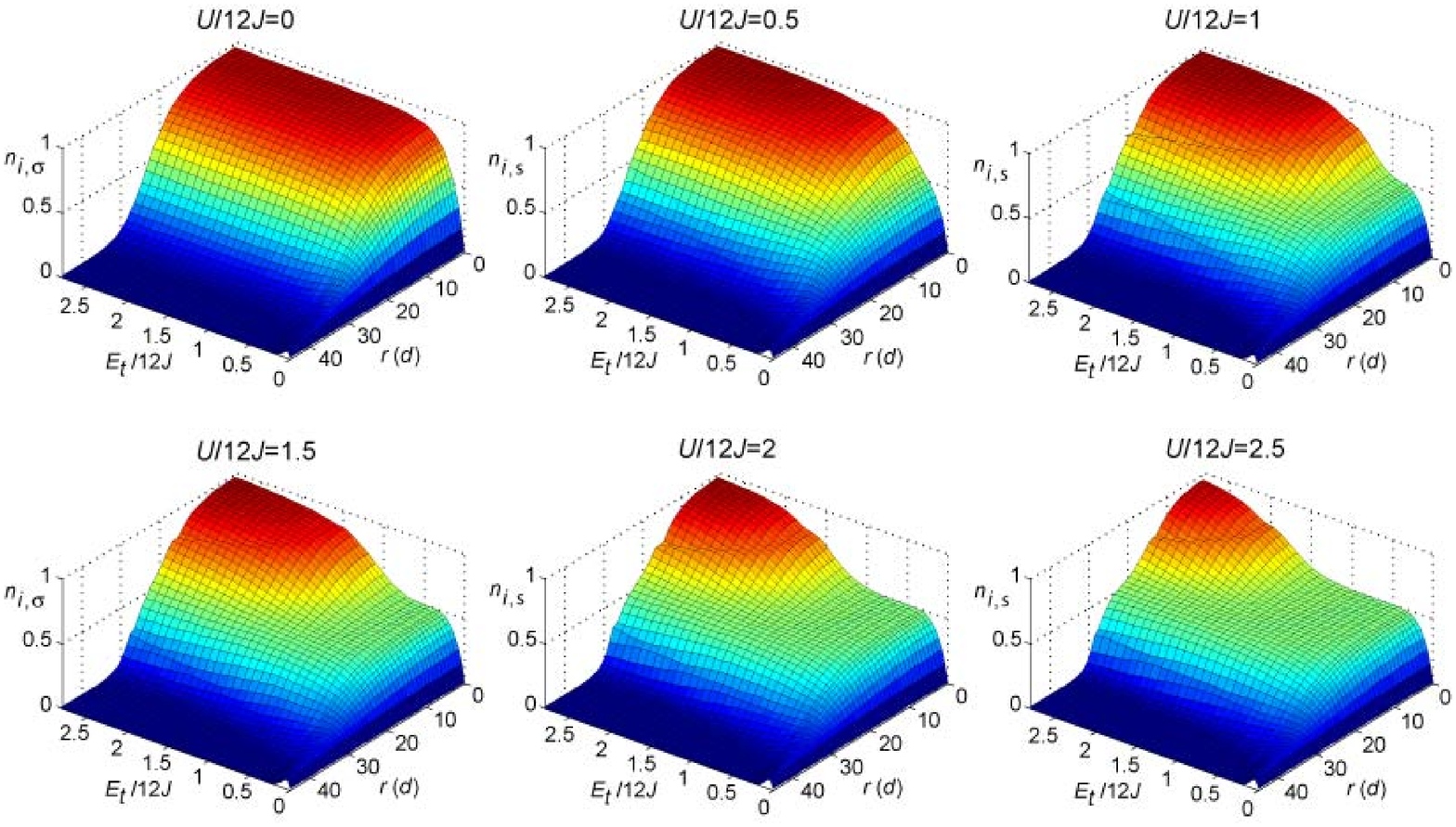}
\caption*{Figure \textbf{S8:} Calculated density distributions for different interactions and an initial temperature of $T/T_F=0.15$. The small ridges for higher interactions are numerical artifacts.
 }
\end{center}
\end{figure}

In the figures S7 and S8 the in-trap density distributions
are plotted for various interaction strengths. The formation of the Mott-insulating core with half filling is clearly visible for both temperatures. \\

\noindent {\bf \Large References and Notes}

\begin{itemize}
\normalsize
\item S1. A. Ino, {\it et al.}, {\it Phys. Rev. Lett.} {\bf 79}, 2101 (1997).
\item S2. S. Ilani, A. Yacoby, D. Mahalu, H. Shtrikman, {\it Science} {\bf 292}, 1354 (2001).
\item S3. R.~W. Helmes, T.~A. Costi. A. Rosch, {\it Phys. Rev. Lett.} {\bf 100}, 056403 (2008).
\item S4. R.~W. Helmes, T.~A. Costi, A. Rosch, {\it Phys. Rev. Lett.} {\bf 101}, 066802 (2008).
\item S5. V. Vuletic, C. Chin, A.~J. Kerman, S. Chu, {\it Phys. Rev. Lett.} {\bf 83}, 943 (1999).
\item S6. P.~S. Julienne, {\it J. Res. Natl. Inst. Stand. Technol.} {\bf 101}, 487 (1996).
\item S7. R. Bulla, T.~A. Costi, D. Vollhardt, {\it Phys. Rev. B} {\bf 64}, 045103 (2001).
\item S8. R. Bulla, T.~A. Costi, T. Pruschke, {\it Rev. Mod. Phys.} {\bf 80}, 395 (2008).
\end{itemize}

\clearpage

\end{document}